# A closer look at the elementary fermions

**Maurice Goldhaber**
Physics Department, Brookhaven National Laboratory, Upton, NY 11973

## Abstract

Although there have been many experimental and theoretical efforts to measure and interpret small deviations from the standard model of particle physics, the gap that the model leaves in understanding why there are only three generations of elementary fermions, with hierarchical masses, has not received the attention it deserves. I present here an attempt to fill this gap. Although our findings are mostly only qualitative, they nevertheless may be of heuristic value. Rules concerning the elementary fermions, some previously known and some new, lead to a number of conclusions and questions that seem worth pursuing. Some clarify the standard model, and others suggest possible modifications, the implications of which are discussed.

## What Can We Learn from the Known Facts About Elementary Fermions?

In many developing fields of science, rules deduced from empirical data often can be considered as qualitative *proto-theories*, having some predictive power as well as pointing the way to a final theory. A well known example is what Mendeleev called his law of the periodic system of the chemical elements, the "elementary particles" of the 19th century. The law, although then far ahead of a plausible theoretical interpretation, nevertheless allowed predictions of new elements. About six decades later it proved helpful in formulating quantum mechanics that in turn explained the periodic system quantitatively, including some of its exceptions. On the way to the final explanation, the periodic system's hierarchical atomic weights were replaced by atomic numbers, equal to the number of protons in the atomic nucleus, with no known *theoretical* limit.

We now face a situation, similar in some respects, for the elementary fermions. A century of research established a "periodic system" of elementary spin 1/2 fermions, confined, however, to *only* three generations. Each generation consists of two kinds of leptons (charged leptons of charge -1 and their associated neutrinos of charge 0) and two kinds of quarks (charges +2/3 and -1/3), with hierarchical masses increasing from one generation to the next.

From the partial width for decay of the neutral gauge boson $Z^0$ of the electro-weak theory into neutrinos, measured ~12 years ago at SLC (SLAC) and LEP (CERN), it was concluded that there are only three kinds of weakly interacting neutrinos, all of low mass (then believed to be zero). If further weakly interacting neutrinos should exist, their masses would have to approach or exceed $1/2\, m_{Z^0}$. Thus, there cannot be more than three generations of elementary fermions with *light* neutrinos. Although the possibility was considered that leptons and quarks might be built of "more fundamental" particles, no indications of complexity (e.g., excited states of the elementary fermions) have been found.

A closer look at the intrinsic properties of the elementary fermions, some measured directly and some deduced with the help of the standard model (SM) of particle physics (1-3), reveals interesting rules about them.

## Some Salient Facts Known for the Three Generations of Elementary Fermions

The SM starts out by postulating chiral symmetries for the interactions of the elementary fermions. Because such symmetries would lead to elementary fermions of zero mass, their finite masses are attributed to symmetry breaking by the so-called Higgs mechanism, which can accommodate masses but not predict them.

Table 1 shows the periodic system of the three generations of elementary fermions with their symbols and masses arranged, as is common, in the order of their discovery. The generations are numbered by a *generation number i*, increasing from 1 to 3 from the lightest to the heaviest generation.

**Table I**
**The Three Generations of Elementary Fermions**

$$i = 3 \quad \begin{pmatrix} t\ (1.743 \pm 0.05) \times 10^5) & \tau\ (1.777 \times 10^3) \\ b\ (4 - 4.3) \times 10^3) & \nu_\tau\ (<2.8 \times 10^{-6}) \end{pmatrix}$$

$$i = 2 \quad \begin{pmatrix} c\ (1.15 - 1.35) \times 10^3) & \mu\ (105.67) \\ s\ (75 - 170) & \nu_\mu\ (<2.8 \times 10^{-6}) \end{pmatrix}$$

$$i = 1 \quad \begin{pmatrix} u\ (1 - 5) & e\ (0.51) \\ d\ (3 - 9) & \nu_e\ (<2.8 \times 10^{-6}) \end{pmatrix}$$

The values of the masses (in MeV) are taken from ref. 4 except for neutrinos, for which only upper limits are known. For $\nu_e$ the limit is based on the work reported recently by the Mainz and Troitsk collaborations (5). Because of the small upper limits for the mass differences *within* the neutrino triplet of mass eigenstates, deduced from neutrino-oscillation experiments (see below), the mass limits for $\nu_\mu$ and $\nu_\tau$ can be taken as approximately the same as that for $\nu_e$.

For the lightest quarks ($u$, $d$, s and c) so-called "current masses" are quoted; they are not measurable directly but are derived from the SM. The masses of complex particles containing these quarks such as, for example, nucleons, are equal to the equivalent of the quarks' potential and kinetic energies. The masses of the heaviest quarks are essentially measured directly.

For elementary fermions in *corresponding* positions in each generation we use the following symbols: within the *i*th generation, we refer to the quarks of charge +2/3 and -1/3, as $u_i$ and $d_i$, respectively, and to the charged leptons and their associated neutrinos as $e_i$ and $\nu_i$, respectively. Our knowledge of the elementary fermions of the first generation stems from experiments in atomic and nuclear physics, whereas that of the second and third generation stems from high-energy experiments, first with cosmic rays, which copiously produce particles of relatively low effective thresholds, and later with high-energy accelerators. Because the chronological order of discovery of the three generations is correlated with the increase of effective collision energies with time, it coincides with the hierarchical order of the masses. The elementary fermions of the first generation, $u$, $d$, and $e$, are widely believed to be the *ultimate building blocks* of which the visible universe is built. They were either produced directly or as final decay products of the quarks and charged leptons of the heavier generations.

Table 2 shows the four known *elementary interactions* (forces) exhibited by the different types of elementary fermions.

**Table II**

**Elementary Interactions of the Elementary Fermions**

| Interactions | Relative Strength | Leptons | | Quarks |
|---|---|---|---|---|
| | | $v_i$ | $e_i$ | $u_i$, $d_i$ |
| Strong | 1 | | | X |
| Electro-magnetic | $10^{-2}$ | | X | X |
| Weak | $10^{-5}$ | X | X | X |
| Gravitational | $10^{-39}$ | X | X | X |

The approximate relative strengths of the interactions, *hierarchically* arranged, are found to be independent of the generation number (universality). Because the interactions vary with energy they are characterized here at a scale of ~ 1 GeV.

By using some experimentally determined properties of the three generations of elementary fermions as input, the SM allowed some important conclusions (see refs. 1-3). For the known interactions the number of quarks and leptons as well as the "flavor" (generation number) of charged leptons are very nearly conserved. The near absence of flavor-changing neutral currents can be explained by the so-called GIM mechanism (named for Glashow, Iliopoulos, and Maiani, ref. 6) as caused by approximate cancellations.

For the Cabbibo-Kobayashi-Maskawa (C-K-M) matrix (7, 8), for which one might have assumed mixing of either $u_i$ or $d_i$ mass eigenstates through their weak interactions, one has chosen to stay with $d_i$ mixing following Cabbibo's original suggestion when only the three quarks *u*, *d*, and *s* were known. The leptonic decays of quarks can then be represented by a unitary matrix:

$$\begin{pmatrix} d' \\ s' \\ b' \end{pmatrix} = \begin{pmatrix} V_{ud} & V_{us} & V_{ub} \\ V_{cd} & V_{cs} & V_{cb} \\ V_{td} & V_{ts} & V_{tb} \end{pmatrix} \begin{pmatrix} d \\ s \\ b \end{pmatrix}$$

The recent review by Gilman, Kleinknecht and Renk (9), gives the following measured ranges for the matrix elements:

$$\begin{pmatrix} 0.9742 \text{ to } 0.9757 & 0.219 \text{ to } 0.226 & 0.002 \text{ to } 0.005 \\ 0.210 \text{ to } 0.225 & 0.9734 \text{ to } 0.9749 & 0.037 \text{ to } 0.043 \\ 0.004 \text{ to } 0.014 & 0.035 \text{ to } 0.043 & 0.9990 \text{ to } 0.9993 \end{pmatrix}$$

We tabulate in Table 3 for different $|i\text{-}j|$ the mean values of the experimental ranges of the matrix elements $|\overline{V}_{u_i d_j}|$. Some matrix elements are derived from more accurately measured ones, assuming a unitary matrix. For each pair $|\overline{V}_{u_i d_j}|$ and $|\overline{V}_{u_j d_i}|$ the ranges partly coincide.

**Table III**

**The Mean Values of the Absolute C-K-M Matrix Elements $|\overline{\nabla}_{u_i d_j}|$**

| $|i\text{-}j|$ | 0 | | | 1 | | | | 2 | |
|---|---|---|---|---|---|---|---|---|---|
| $i$ | 1 | 2 | 3 | 1 | 2 | 2 | 3 | 1 | 3 |
| $j$ | 1 | 2 | 3 | 2 | 1 | 3 | 2 | 3 | 1 |
|   | 0.9750 | 0.9742 | 0.9992 | 0.223 | 0.223 | 0.040 | 0.039 | 0.004 | 0.009 |

From Tables 1-3 we can deduce the following rules.

**Rule 1.** Corresponding elementary fermions of different generations are associated with identical elementary interactions (universality).

**Rule 2.** Within each generation, there is a correlation between the mass of an elementary fermion and the relative strength of its dominant interaction (elaborated further later).

**Rule 3.** Besides its dominant interaction, each elementary fermion possesses *all* weaker ones.

**Rule 4.** As the generation number $i$ increases, the mass differences $m(u_i) - m(d_i)$ and $m(e_i) - m(v_i)$ also increase.

**Rule 5.** The matrix elements of the C-K-M matrix decrease as $|i\text{-}j|$ increases from 0 to 2, with $|\overline{\nabla}_{u_i d_j}| \approx |\overline{\nabla}_{u_j d_i}|$.

According to the SM, each elementary fermion may emit or absorb elementary gauge bosons connected with its elementary interactions, either virtually or really, depending on energy.

## What Can the Rules Teach Us?

Rule 2 suggests that self-interactions may be responsible for a major part or all of the mass of an elementary fermion.

Extrapolations from rules may have predictive value. There are many examples in the history of science where an extrapolation or generalization from existing knowledge was taken as a prediction, often only qualitative but worth pursuing, and in many cases resulting in important progress. This was especially so in the early steps leading to the SM. If one wants to be more cautious, however, one might consider extrapolations as questions worth pursuing until they are either confirmed experimentally or integrated into theory, not necessarily in that order.

Rules 1-3 invite the intriguing question: Are there additional members of each generation, elementary spin 1/2 fermions, subject *solely* to the gravitational interaction, and thus expected to have extremely small masses? Can such *gravity fermions* be integrated into the theory of general relativity without running into inconsistencies?

Rules 1 and 2 imply that *within* each generation the order of the masses of the elementary fermions is correlated with the hierarchy of their *dominant* self-interactions.

Rule 4 indicates that the $u_i$ and $d_i$ (except for the special case of the first generation discussed later) have large mass differences, contrary to expectations from quantum chromodynamics (QCD). Should we therefore treat them, in analogy with the $e_i$ and $v_i$, as elementary fermions with different interactions, implying that a new dominant *hyperstrong* elementary interaction may have to be added to the QCD interaction for the $u_i$?

Rule 2 is consistent with the existence of finite neutrino masses as deduced from recent neutrino oscillation experiments. This rule also allows us to sharpen and extend the conclusion drawn from the $Z^0$ experiments; because the dominant self-interaction of neutrinos is the weak

5one, we can expect *only* low-mass neutrinos. If a new neutral elementary particle of large mass and spin 1/2 should be discovered, e.g., as a component of cold dark matter, it would have to be considered *sui generis*.

Rule 5 suggests that the amount of weak mixing of quarks depends on the "closeness" of the generations involved. In his different parameterization of the C-K-M matrix, Wolfenstein already emphasized the hierarchical reduction in the extent of mixing (see refs. 9 and 10).

From a particular version of the superstring theory, Candelas *et al.* (11) predicted the existence of four generations of elementary fermions. However, Erler and Langacker (see ref. 4) concluded from a review of several precision measurements of quantities sensitive to virtual effects of elementary fermions of a hypothetical fourth generation that there is no evidence for it even if its members were too heavy to have been detected directly at presently available energies.

Many theoretical approaches attempt to understand some particular empirical input to the SM or to reproduce some of its results in a new way. These approaches usually are based on special assumptions going beyond the SM and not on a comprehensive, generally accepted theory. For two interesting examples, see refs. 12 and 13.

## Attempts to Understand the Empirical Regularities

Several questions remain: What do the rules imply, what causes the hierarchy of the masses of corresponding members of *different* generations, and why are there only three generations of elementary fermions?

The SM does not have satisfactory answers to these questions; however, the empirical data suggest modifications of the SM that may qualitatively explain the observations. Although the SM assumes no *a priori* difference between the different generations, we conclude from rules 4 and 5 that the elementary fermions "know" to which generation they belong and that the generation number is not just a label, as is usually assumed, but stands for a *new physical property* equal for all members of a particular generation but changing systematically from one generation to the next. It is apparently a property that *enhances* self-interaction, more so the larger $i$ becomes. What can such a physical property be? While retaining the universality of the interactions, for which there is good empirical evidence, a possible tentative interpretation is to forgo the assumption of *equal point sources* for the three generations and replace them with *source shapes* of finite size, identical for each type of elementary interaction *within* a generation but decreasing systematically in volume as $i$ increases; as the source volume becomes more "singular," the self-interactions will increase. The masses of the elementary fermions then have to be considered as *secondary* quantities that take on hierarchical values.

When we find nature repeating itself with variations, it is worth asking: Why should the repetition stop at three? I conjectured some time ago that the volume of the source shapes might decrease "naturally" if the existence of just three generations were connected with the three-dimensionality of space (14). The source shapes of the three generations might resemble, for example, a sphere, a disk, and a rod (with no zero thickness and no sharp edges); as $i$ increases from 1 to 3, their dimensionality decreases from 3 to 1, and their shapes become more singular.

Although the value of the masses of the elementary fermions is correlated with the strength of their dominant self-interaction, nondominant self-interactions also must play a role, as they do when *they* are the dominant self-interactions in lighter elementary fermions of the same generation. The effect of a nondominant self-interaction on the mass may be positive or negative depending on whether it has the same or opposite sign from that of the dominant interaction. This effect may be especially important for the strong and electromagnetic self-interactions, which are near in relative strength and presumably have opposite signs (the dominant interaction is attractive and the nondominant one is repulsive, *independent* of the sign of the electric charge), leading to a net reduction in the self-interaction. Compared with QCD expectations alone, $u_i$ and $d_i$ both will be depressed but $u_i$ more than $d_i$. In the first generation, this might be the cause of the lower $u$ than $d$ mass, which would explain the long-standing puzzle of why the neutron (containing one $u$ and two $d$) is heavier than the proton (containing two $u$ and one $d$). But then one may ask: why is $u_i$ not lower than $d_i$ for all $i$ (see later discussion)?

With the discovery of oscillations of atmospheric neutrinos $\nu_\mu \to \nu_\tau$, near-maximal mixing of the neutrino mass eigenstates $m_2$ and $m_3$ was established (for the latest results, see refs. 15 and 16). Of several potential oscillation solutions still under consideration for solar neutrinos, a large mixing angle solution is preferred (see ref. 17), compatible with measurements of the solar $^8$B neutrino spectrum by SuperKamiokande (18, 19) and the Sudbury Neutrino Observatory (20).

From the oscillation experiments which yield a value for $|m_3^2 - m_2^2|$ of $(2 - 5) \times 10^{-3}$ $(eV)^2$ (16, 17) and for $|m_2^2 - m_1^2| < 7 \times 10^{-4}$ $(eV)^2$ (21), assuming maximal mixing, a spread of $<0.1$ eV for the mass eigenstates of the neutrino triplet was deduced (22). The near degeneracy of the mass eigenstates makes large mixing plausible. The often-conjectured hierarchical order of the neutrino masses might be expected from the source shapes assigned to the three generations.

## Epilogue

We have seen that the known properties of the elementary fermions lead to rules through which nature *speaks to us*, suggesting clarifications and modifications of some assumptions of the SM as well as hinting at possible new avenues worth exploring.

Some comments may be useful. We learned that elementary interactions are connected with appropriate elementary fermions from each generation and raised the question of whether this should be extended to the gravitational interaction.

Although the assignments of corresponding elementary fermions to different generations in hierarchical order of their masses, including leptons and quarks that have no other known associations, can be ascribed to the happenstance of the chronological order of their discovery, we have made it plausible that this is to be expected for the assumed source shapes.

Although many successes of the SM were not discussed here and usually will not be affected by our considerations, small deviations from its predictions may result from some of the suggested modifications.

For finite source shapes, the dependence of the C-K-M matrix elements on the generation distance (Rule 5) can be expected to depend on the amount of overlap between the source shapes of the initial and final quarks. This allows an alternative interpretation to $d_i$ mixing; if the "intrinsic" charge current weak interaction was assumed to be *independent* of generation change, the matrix elements, as long as the off-diagonal elements are small, would be a measure of the overlap of the source shapes, and the unitarity limit would be expected to be approached.

The large increases in the mass difference between $u_i$ and $d_i$ with $i$ (rule 4) indicate that the postulated hyperstrong interaction of the $u_i$ would have more impact on the self-interaction as the shapes become more singular. This would be so if it is mainly a *short-range high-energy* interaction, whereas the effective QCD self-interaction is a comparatively *low-energy* interaction at short distance. The assumed source shapes may cause the $u_i$ masses to be lifted well above the $d_i$ masses for $i > 1$. For the spherical source shape assumed for $i = 1$, the $u$ mass would be less affected by a high-energy interaction. The relative closeness of the $d_i$ to the $e_i$ despite the large difference in their dominant interaction strengths may be caused in part by differences in the energy dependence of their interactions and in part by the depression of the $d_i$ by their Coulomb interaction.

If a hyperstrong interaction exists, one would expect that the cross sections for producing t̄ + t at energies that are well above their threshold would deviate from QCD predictions.

If a gauge boson is connected with the hyperstrong interaction, it might be detectable by its characteristic mass, decay modes, and lifetime.

Bardeen *et al.* (23 and references therein) discuss dynamical symmetry breaking of the SM by a t̄t condensate where a new interaction (less general than the hyper-strong interaction proposed here), called topcolor (24 and references therein), is ascribed to the t. The Higgs boson is then considered to be a t̄t condensate, with a mass <500 GeV, estimated by Chivukula (25).

The existence of a hyperstrong interaction also would affect the estimate of the grand unification mass and thus the predictions of Grand Unified Theories for the proton lifetime.

If elementary fermions with only gravitational interactions exist, the "gravity fermions" might be a link between general relativity and quantum mechanics. They also would contribute to hot dark matter (as would spin 2 gravitons that often are ignored in this connection). The gravitational wave detectors now being built are only sensitive to coherent waves of gravitons and would not detect individual gravity fermions that might only be detectable through interactions at extremely high energies, of the order of the Planck mass ($10^{19}$ GeV), but such high energies have not been found among the cosmic rays.

Fermions of spin 3/2, with gravitational interactions only, are predicted by the theory of supergravity (26).

Ongoing improved experimental and theoretical studies of the values of the C-K-M matrix elements, corrected, in the case of bound quarks for the different wave functions of the initial and final quarks, may make the *a priori* assumption of unitarity unnecessary.

Although our considerations seem to yield a coherent picture, with important parts independent of the assumption of finite source shapes, clarifying some aspects of the SM and raising the possibility of modifying and extending it, a crucial question remains: Can our qualitative approach be changed into a quantitative one without running into contradictions? Can one find source shapes (compatible in effective size with the present experimental limits of $\sim 10^{-17}$ cm) that, unlike point sources, would yield finite masses and lead, for the well established self-interactions, to the masses found for the elementary fermions (including the mass difference *d-u*) e.g., by inversion and iteration, and would such source shapes also explain the C-K-M matrix elements?

## Acknowledgements

I thank M. J. Creutz, M. V. Diwan, A. S. Goldhaber, T. Goldman, R. L. Jaffe, W. J. Marciano, D. J. Millener, and R. E. Shrock for valuable discussions.

## Note

My attention was drawn to three attempts to understand various aspects of the three generations of elementary fermions. They differ from each other and from the views presented here (27-29).